\documentclass[twocolumn, showpacs,preprintnumbers,amsmath,amssymb]{revtex4}

\usepackage{amsmath,amsfonts,amssymb}
 \usepackage{epsf}

\newcommand{\be}{\begin{equation}}
\newcommand{\ee}{\end{equation}}

\newcommand{\bea}{\begin{eqnarray}}
\newcommand{\eea}{\end{eqnarray}}

\newcommand{\bra}{{\langle}}
\newcommand{\ket}{{\rangle}}
\newcommand{\tr}{\hbox{ Tr}}

 \newcommand{\myfig}[3]{\begin{figure}[ht]
\begin{center}
\leavevmode \epsfxsize=#2cm \epsfbox{#1}
\end{center}
\caption{#3} \label{fig:#1}
\end{figure}}

\begin{document}

\preprint{}

\title{Reconstructing 1/2 BPS Space-Time Metrics from Matrix Models
and Spin Chains}

\author{Samuel E. V\'azquez}
 \email{svazquez@physics.ucsb.edu}
 \affiliation{Department of Physics, UCSB, Santa Barbara, CA 93106 }
\date{}

\begin{abstract}
Using the AdS/CFT correspondence, we address the question of how to
measure complicated space-time metrics using gauge theory probes. In
particular, we consider the case of the 1/2 BPS geometries of type
IIB supergravity. These geometries are classified by certain
``droplets" in a two dimensional space-like hypersurface. We show
how to reconstruct the full metric inside these droplets using the
one-loop ${\cal N} = 4$ SYM theory dilatation operator. This is done
by considering long operators in the $SU(2)$ sector, which are dual
to fast rotating strings on the droplets. We develop new powerful
techniques for large $N$ complex matrix models  that allow us to
 construct the Hamiltonian for these strings. We find
that the Hamiltonian  can be mapped to a ``dynamical" spin chain.
That is, the length of the chain is not fixed. Moreover, all of
these spin chains can be explicitly constructed using an interesting
algebra which is derived from the matrix model. Our techniques work
for general droplet configurations. As an example, we study a single
elliptical droplet and the ``Hypotrochoid".

\end{abstract}

\pacs{Valid PACS appear here}

 \maketitle

\section{Introduction}

One of the most striking predictions of the AdS/CFT correspondence,
is the emergence of space-time geometry from the large $N$ limit of
non-abelian gauge  theories. The best understood example  is the
duality between ${\cal N} = 4$ SYM theory on $\mathbb{R}\times S^3$
and type IIB String Theory on asymptotically $AdS_5\times S^5$
space-times \cite{malda}.

It was understood early on, that the ground state of ${\cal N} = 4$
SYM was dual to $AdS_5\times S^5$ itself. This matching was
originally guessed based on the symmetries of the ground state.
However, it was latter understood that one can also consider states
dual to fast rotating strings on this background \cite{BMN}. These
states allowed a matching of the string spectrum, including the
sigma model,  in the appropriate limits \cite{Kruczenski}. (See
\cite{Tseytlinrev} for a review and more references.) By matching
the sigma-model of these strings one is also measuring the
space-time metric of $AdS_5\times S^5$.

More recently, it has been possible to identify  space-times that
are dual to ``heavy" 1/2 BPS states of the gauge theory \cite{LLM}.
These geometries have complicated metrics and topologies.
Nevertheless, it was shown in \cite{LLM} that they are classified in
terms of ``droplets" in a plane.  Remarkably, the 1/2 BPS states of
SYM are classified in exactly the same way \cite{toy,matrixmaps}.

Space-times corresponding to 1/4 and 1/8 BPS states have also been
studied in \cite{148bps}. Their classification in terms of gauge
theory operators is not completely understood, but some proposals
have been put forward \cite{davidN,Ryzhov}.

In this paper, we address the question of how to measure the
complicated metrics of the 1/2 BPS geometries using only gauge
theory probes. To simplify the problem we focus on the $SU(2)$
sector of the gauge theory. As we explain below, this sector
correspond to strings that live inside the ``droplets" and rotate
along an $S^1$ fiber. As usual, one can match a one-loop gauge
theory calculation by studying strings with large angular momentum
along this circle.

On the gauge theory side, the $SU(2)$ dilatation operator can be
described by  a model of matrix quantum mechanics with two matrices
\cite{beisert}. Using this model, we explain how one can excite a
``heavy" 1/2 BPS state, and then put a probe string on it. The
reduced Hamiltonian for the probe string can be computed using
Random Matrix Theory techniques. Some of these techniques are
developed here for the first time.

We find that the Hamiltonian of the probe string can be described by
a bosonic lattice. This model has the usual hopping terms, but also
include sources and sinks of bosons at each site. Alternatively, one
can visualize the lattice as a ``dynamical" spin chain. As it turns
out, this Hamiltonian is completely determined by a very interesting
algebra underlying the Random Matrix Model. Using coherent states,
one can match the thermodynamic limit of this lattice to the sigma
model of the fast string. This is how we measure the metric.

 The techniques developed here are valid for  any droplet
 configuration, including the case of multiple-connected domains.
 As an example, we study a single  elliptical droplet, and the so-called ``Hypotrochoid".

 The paper is organized as follows. In section II we review the 1/2
 BPS geometries and set up the basic notation. In section III, we
 study the fast rotating strings.  We derive their sigma-model for a
 general droplet configuration. We then specialize to the case of
 a single elliptical droplet, and the Hypotrochoid. In section IV we
 derive some general results for the gauge theory. We start with a review of the
 $SU(2)$ sector in order to set up notation. We then define the
 Hilbert Space for a probe string  around an arbitrary
 1/2 BPS background.  Next, we derive the corresponding
 Hamiltonian using the one-loop dilatation operator. Coherent states
 are defined, and  an effective sigma model is derived. At this
 stage, we show that the general form of the sigma model is exactly
 the same as the one found in the String Theory calculation. This
 allows the full metric on the droplet to be defined in terms of gauge theory
 quantities.

 In sections V and VI, we consider the particular examples of the
 elliptical droplet and the Hypotrochoid, this time from the gauge
 theory side. We reproduce all String Theory results.

 We conclude in section VI with a discussion which includes topics
 such as integrability, the prospect of probing Black Hole
 states, and extension of this procedure to other sectors of the
 gauge theory.

 The reader familiar with Normal Random Matrices will find that this
 paper is basically an extension of these techniques to large $N$
 {\it complex} ensembles.

\section{1/2 BPS Metrics}
All 1/2 BPS solutions to type IIB supergravity with $N$ units of RR
five-form flux have been found in \cite{LLM}. They preserve a
bosonic $\mathbb{R} \times SO(4) \times SO(4)$ symmetry of the ten
dimensional space-time.  All solutions are classified by a single
function, which we  call $\rho$, on a two dimensional plane. The
metrics can be written as, \bea \label{metric}ds^2 &=& -h^{-2}(Dt)^2
 + h^2 (dy^2 + dz d\bar z) + y e^{-G} d\Omega_3^2\nonumber \\
 && +  y e^{G}
 d \widetilde{\Omega}_3^2 , \nonumber \\
 h^{-2} &=& 2 y \cosh G,\nonumber \\
 f &=& \frac{1}{2} \tanh G, \nonumber \\
 f(z,\bar z, y) &=&   - \frac{y^2}{2} \int d^2 z' \frac{\rho(z',\bar z')}{(|z -
 z'|^2 + y^2)^2}.
 \eea
Here, we have defined the covariant derivative, $Dt = dt + V = dt  -
\frac{1}{2} i \bar V dz + \frac{1}{2} i  {V} d\bar z$, and we are
using complex coordinates in the $y = 0$ plane: $z = x_1 + i x_2,
\bar z = x_1 - i x_2$. Moreover, \be V(z,\bar z, y) = -V_2 + i V_1 =
\frac{1}{2} \int d^{2}z' \frac{\rho(z',\bar z')(z - z')}{(|z - z'|^2
+ y^2)^2}\;.\ee

All non-singular solutions must have $\rho = \pm 1/\pi$. Therefore
we can separate the integrations above in domains or ``droplets"
(${\cal D}_i$) for which $\rho = 1/\pi$ (say) inside and $\rho =
-1/\pi$ outside. These are the configurations that we will consider
in this paper.

The size of the asymptotic $AdS_5 \times S^5$ is set by \cite{LLM},
\be R_{AdS_5}^4 = R_{S^5}^4 = \sum_i \int_{{\cal D}_i} d^2 z
\rho(z,\bar z)\;.\ee Therefore, we can rescale all {\it spatial}
coordinates by $x^i \rightarrow R^2_{AdS} x^i$ and we get an overall
factor of $R_{AdS}^2$ in front of the metric. Then, our area
quantization condition is simply, $\sum_i \int_{{\cal D}_i} d^2z
\rho = 1$. This makes it easier to compare with the gauge theory.

For a single simply connected domain $\cal D$, we can rewrite the
integrals over the droplet as contour integrals over its boundary.
We have, \bea  V &=& \frac{1}{2} \int_{\cal D} \frac{d^2 z'}{\pi}
\bar
\partial' \frac{1}{|z - z'|^2 + y^2} \nonumber \\
&&- \frac{1}{2} \int_{\mathbb{C} \backslash {\cal D}} \frac{d^2
z'}{\pi} \bar
\partial' \frac{1}{|z - z'|^2 + y^2} \nonumber \\
&=&  \oint_{\partial D} \frac{dz'}{2\pi i} \frac{1}{|z - z'|^2 +
y^2}\;. \eea Similarly, \bea f &=& -\frac{1}{2} \int_{\cal D}
\frac{d^2 z'}{\pi} \bar\partial' \partial' \log\left(|z - z'|^2 +
y^2\right)\nonumber \\
&&+  \frac{1}{2} \int_{\mathbb{C} \backslash \cal D} \frac{d^2
z'}{\pi} \bar\partial'
\partial' \log\left(|z - z'|^2 + y^2\right)\nonumber\\ &=&
\frac{1}{2} -  \oint_{\partial D} \frac{dz'}{2\pi i} \frac{\bar z' -
\bar z}{|z - z'|^2 + y^2} \;.\eea  The factor of $1/2$ in $f$ comes
from the boundary at infinity, and the contour integrals are taken
counterclockwise.

This procedure is valid even for multiple droplets or non-simply
connected domains. One obtains a superposition of contour integrals
over the different boundaries of the domains. The orientation of the
contours is such that we keep the regions with $\rho = 1/\pi$ (the
area of ${\cal D}_i$) to our left. These contour integrals can be
solved, in general, by a conformal map from the boundaries to the
unit circle. The problem then reduces to holomorphic contour
integrals. We will illustrate this in the next section.

\section{Semiclassical Strings in the $SU(2)$ Sector}
The $SO(4)$ subgroups in the isometry group of the 1/2 BPS
geometries, correspond to rotations on the two $S^3$s that appear in
the metric (\ref{metric}). In particular, one can see that in the
asymptotic region, $\widetilde{S}^3 \subset AdS_5$ and $S^3 \subset
S^5$. Inside each droplet ($y = 0$), the size of $\widetilde{S}^3$
is zero.

The $SU(2)$ sector of the gauge theory consist of states with two
R-charges. These two charges are identified (asymptotically) by a
rotation along $S^1 \subset S^3$, and a rotation around the origin
in the $y = 0$ plane. Therefore, one expects that the strings dual
to these operators will ``live" inside each droplet, and will rotate
along an $S^1$ fiber.

The metric restricted to this subspace can be calculated by setting
$y \rightarrow 0$ in (\ref{metric}) and restricting $z,\bar z$
inside the droplet. A careful calculation shows that, \be
\label{su2metric} \left. ds^2\right|_{SU(2)} = -h^{-2} (Dt)^2 + h^2
dz d\bar z + h^{-2}d\varphi^2\;,\ee where, for a single droplet,
\bea\label{h} h^{4} &=& \oint_{\partial D} \frac{dz'}{2\pi i}
\frac{\bar z' - \bar z}{|z - z'|^4}\;, \\ \label{V} V &=&
\oint_{\partial D} \frac{dz'}{2\pi i} \frac{1}{|z - z'|^2}\;.\eea

This result can  be  generalized to the case of multiple droplets
and for non-simply connected domains. One just needs to superimpose
integrations over the different boundaries with the appropriate
orientations. The important point to notice is that, in general, we
can write, \be \label{Kahler} V = \bar \partial \log K\;, \;\;\; h^4
=
\partial V = \partial \bar
\partial \log K\;,\ee where we define, \be \label{Kpot}\log K \equiv \sum_i
\oint_{\partial D_i} \frac{dz'}{2\pi i} \frac{1}{z - z'}\log(\bar z
- \bar z')\;.\ee As we will see, the ``K\"{a}hler potential" $K$
will have a very special interpretation as a sum of orthogonal
polynomials.

Let us now look at the fast string limit along $\varphi$. As usual
\cite{kru}, we start with the Polyakov action in momentum space,
\begin{eqnarray}
\label{paction} S_p = \sqrt{\lambda_{YM}} \int d\tau \int_0^{2\pi}
\frac{d\sigma}{2 \pi} {\cal L}\;,
\end{eqnarray}
where,
\begin{eqnarray*}
{\cal L} &=& p_\mu \partial_0 x^\mu + \frac{1}{2} A^{-1} \left[
G^{\mu \nu} p_\mu p_\nu + G_{\mu \nu} \partial_1 x^\mu
\partial_1 x^\nu \right] \nonumber \\
&&+ B A^{-1} p_\mu \partial_1 x^\mu\;.
\end{eqnarray*}
Remember that we have factorized the radius of $AdS$ and so that, by
the $AdS$/CFT correspondence,  $\lambda_{YM} = g_{YM}^2 N  =
R_{AdS}^4/\alpha'^2 \equiv 8 \pi^2 \lambda$. Moreover,  $A$, $B$
play the role of Lagrange multipliers implementing the Virasoro
constraints $T_{a b} = 0$.

As usual, the natural gauge choice is the one that distributes the
angular momentum uniformly along $\varphi$.  Thus our gauge is, \be
t = \tau\;,\;\;\; p_\varphi = \text{const.}\ee The angular momentum
along the $\varphi$ coordinate is,
\begin{eqnarray*}
 L  = \sqrt{\lambda_{YM}} \int_0^{2\pi} \frac{d\sigma}{2 \pi}
p_\varphi = \sqrt{\lambda_{YM}} p_\varphi \;.
\end{eqnarray*}

The Virasoro constraints for our metric (\ref{su2metric}) read, \bea
\label{virasoro1} -h^2 p_t^2 + h^{-2} (|p|^2 - V_1^2) + h^2
p_\varphi^2 + h^2 |z'|^2 + h^{-2}
\varphi'^2 &=& 0 \;,\nonumber \\\\
\label{virasoro2}p_t V_1 + p z' + \bar{p} \bar{z}' + p_\varphi
\varphi' &=& 0 \;.\nonumber \\ \eea The notation for the one-form
$V$ is the following: \be V_a =  V_\mu \partial_a x^\mu =
\frac{1}{2} i  {V} \partial_a \bar z - \frac{1}{2} i \bar V
\partial_a z\;.\\
 \ee

 The procedure now is the same as in \cite{kru}. First, we solve
 for $p_t$ in terms of the spatial momenta using the Virasoro
 constraints (\ref{virasoro1}) and (\ref{virasoro2}). We then plug this into the action
 (\ref{paction}), so that the resulting action depends on
 $(x^i,\dot{x}^i, x'^i, p_i)$ only, where $i = z, \bar z$. Finally, since the momenta $p_i$ will only
 enter the Lagrangian algebraically, we can solve for them in terms
 of $(x^i,\dot{x}^i, x'^i)$ and plug the result back into the action. At
 this point, one considers the limit where $p_\varphi >> 1$.
 Moreover, one assumes that the fields are slowly varying in time so
 that $\dot{x}^i \sim {\cal O}(1/p_\varphi^2)$. With this
 assumption, one then follows the procedure of \cite{kru} to eliminate
 higher powers of the time derivatives in terms of higher orders of
 the spatial derivatives. This can be done systematically, and one
 finds an action that is linear in the time derivatives. This form
 of the action is more natural when comparing with the gauge theory
 calculations.

For the leading order in $1/p_\varphi^2$, we do not need to
 follow this complicated procedure. Just expanding to ${\cal
 O}(1/p_\varphi^2)$ and with the assumptions above, we find,
 \bea \label{redsigma}
S_p &\approx&  -  L \int d\tau \int_0^1 d\sigma \left[  \frac{i}{2}
V \dot{\bar{z}}  - \frac{i}{2} \bar{V} \dot{z} + 1 +
\frac{\lambda}{L^2} |z'|^2 \right. \nonumber \\
&&\left. + {\cal O}\left(\frac{\lambda^2}{L^4}\right)\right]\;.\eea

Therefore, we see that at one loop, the effective string action has
a very universal form. Nevertheless it does incorporates non-trivial
aspects of the underlying geometry. Note that the canonical momenta
to the coordinates $z$ and $\bar z$ are determined by the one-form
$V$. We will latter see how this one-form arises from the gauge
theory. The non-triviality of this one-form reflects the fact that
our system has the constraint: $z \in {\cal D}$. One can, in
practice go to higher loops and start seeing the emergence of the
function $h$. The same is possible in the gauge theory side.
However, since the full metric (\ref{su2metric}) is completely
determined by the ``K\"{a}hler potential" (\ref{Kpot}), we will not
pursue this here.

\subsection{Calculation of the one-form $V$}
In this section we show how to calculate $V$ for a general droplet.
We will show explicit results for the elliptical droplet and the
Hypotrochoid. The procedure presented here, is also applicable for
multiple droplets and non-simply connected domains.  In fact, it can
also be used to calculate $V$ and $h$ outside  the plane of the
droplet.

One starts by noting that the exterior of the droplet
$\mathbb{C}\backslash {\cal D}$ can be conformally  and univalently
mapped to the exterior of the unit disk \cite{matrixreview, growth}.
We call this map $w(x)$ and its inverse $x(w)$. These maps are
illustrated in figure 1. \myfig{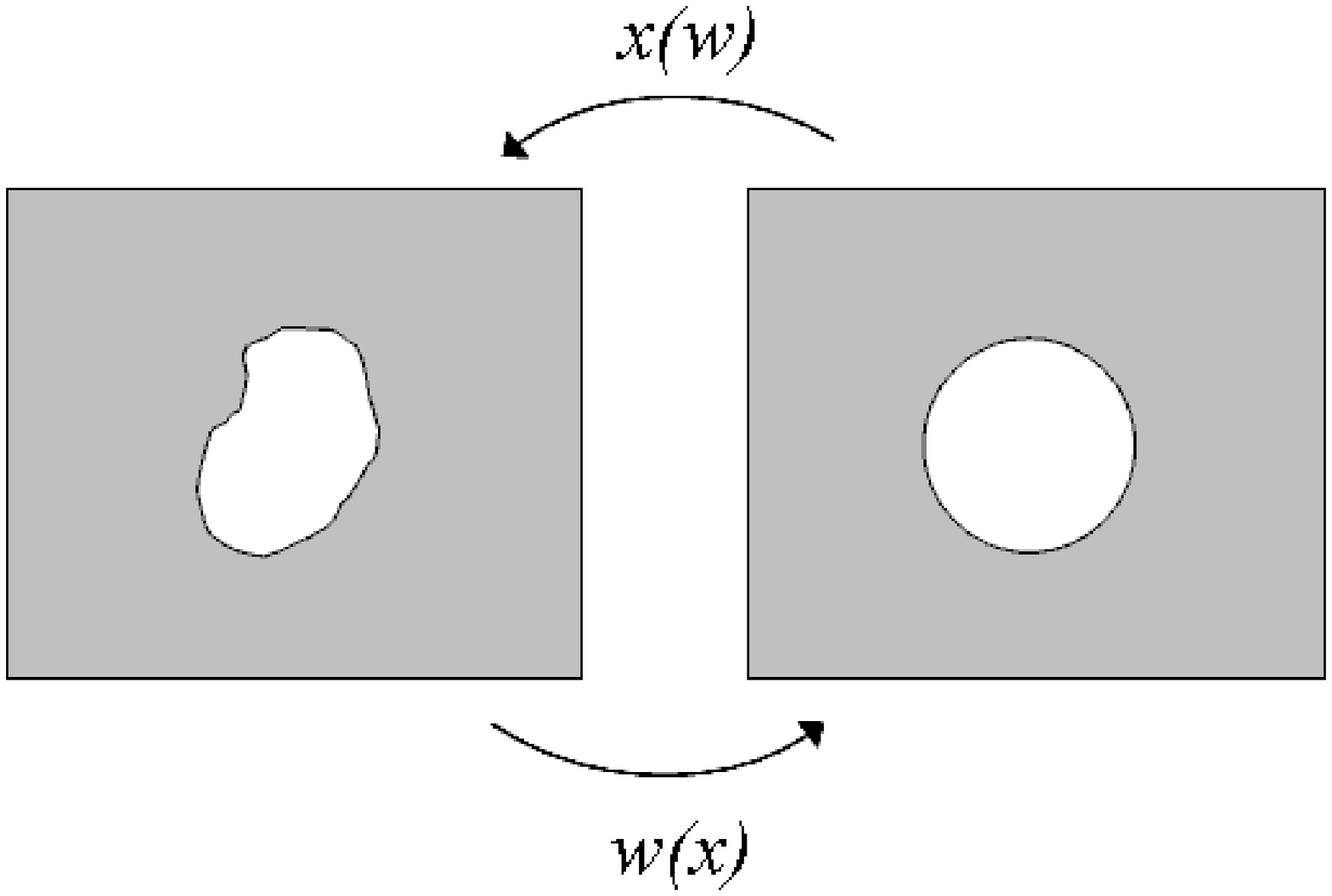}{7}{The conformal map
$w(x)$ and its inverse $x(w)$. }

The inverse conformal map takes the form of a semi-infinite Laurent
polynomial, \be\label{invconf} x(w) = r w + \sum_{k>0} u_k w^{-k}\;,
\;\;\; |w|
> 1\;.\ee The coefficient $r$ can be chosen to be real and positive.
Our interest is the boundary of the domain. Therefore, we can take
the limit where $w$ approaches the boundary of the unit disk: $w
\rightarrow e^{i \phi}$. In this case we obtain the map, $x: S^1
\rightarrow
\partial {\cal D}$. In general, this map can be multi-valued, so we
need to be careful in choosing the parameters of the conformal map
to avoid this.

Once we have the inverse conformal map (\ref{invconf}), we can
convert any contour integral over $\partial{\cal D}$ to a
holomorphic contour integral over $S^1$. For example, Eq. (\ref{V})
becomes \be\label{Vint} V(z,\bar z) = \oint_{S^1} \frac{dw}{2\pi
i}x'(w) \frac{1}{(x(w) - z)(\bar{x}(w^{-1}) - \bar{z})}\;,\ee where
the prime is the derivative with respect to $w$. The value of the
integral will then be the sum of the residues of the simple poles in
$w$ that are inside the unit circle.

Let us now note, that not all the parameters in the conformal map
are independent. In our case, we need to impose the restriction on
the area of the droplet. This is given by, \bea\label{areanorm} 1
&=& \int_{\cal D} \frac{d^2z}{\pi}  = \int_{\cal D} \frac{d^2z}{\pi}
\bar  \partial \bar z = \oint_{ S^1} \frac{dw}{2\pi i} x'(w) \bar
x(w^{-1})\;.\nonumber \\ \eea

It is very convenient to write the parameters of the inverse
conformal map, in terms of the so-called ``moments" of the domain
${\cal D}$.  Assuming that we include the origin in the droplet, the
moments are defined by ($k  > 0$), \be \label{moments} t_k =
\frac{1}{ k} \oint_{\partial \cal D} \frac{dz}{2\pi i} \bar z z^{-k}
= \frac{1}{k} \oint_{S^{1}} \frac{dw}{2\pi i} \frac{x'(w)
\bar{x}(w^{-1})}{ x^k(w)}\;.\ee We will see that these moments
translate directly to the parameters of the gauge theory operators.

\subsection{Elliptical Droplet}
For the elliptical droplet, we can use the simplest singular inverse
map, $x(w) = r w + \frac{u}{w}$. Using the area normalization
(\ref{areanorm}), we get, $1 = r^2 - |u|^2$. Without lost of
generality we can take $u$ to be real \footnote{A phase in $u$ will
correspond to a rotation of the droplet.}. Using (\ref{moments}),
one can see that the elliptical droplet has only $t_2 \neq 0$. The
 inverse map can then be parameterized by, \be x(w) =
\frac{1}{\sqrt{1 - \epsilon^2}} \left(w +
\frac{\epsilon}{w}\right)\;, \;\;\; 0 \leq \epsilon < 1\;,\ee where
$\epsilon \equiv 2 t_2$ is the eccentricity of the ellipse. The
 elliptical droplet is shown in figure 2.
 \myfig{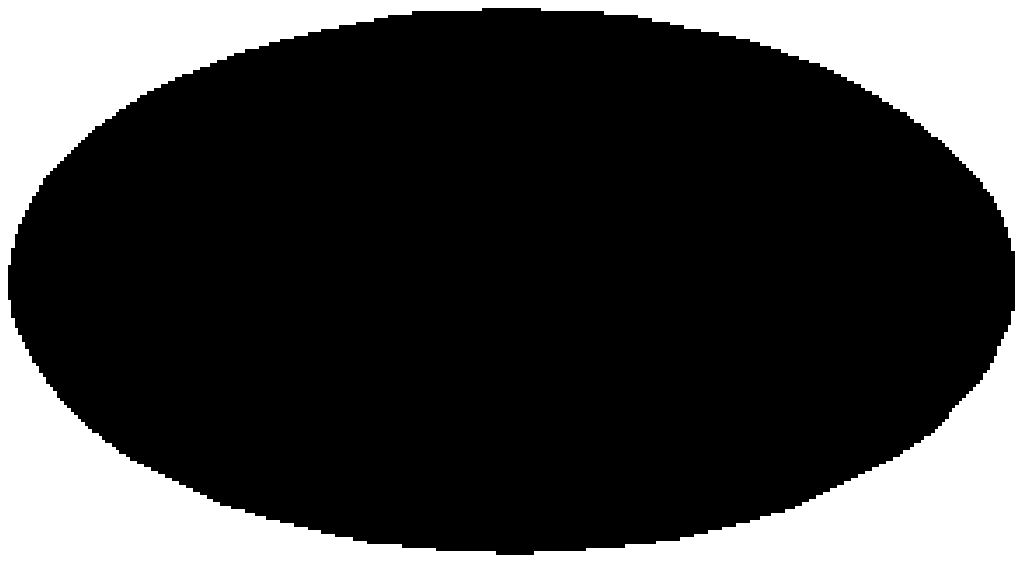}{4}{Elliptical Droplet. We have chosen $t_2$ to be real.}

The integrand in (\ref{Vint}) has the form: \be\frac{\sqrt{1 -
\epsilon^2}}{\epsilon} \frac{w^2 - \epsilon}{(w - w_1)(w - w_2)(w -
\bar w_1/\epsilon)(w - \bar w_2/\epsilon)}\;,\ee where the poles are
given by, \be w_{1, 2}=\frac{1}{2}\sqrt{1 - \epsilon^2}\left( z \pm
\sqrt{z^2 - \frac{4\epsilon}{1 - \epsilon^2}}\right)\;.\ee

Now we need to find out which poles are inside the unit circle. To
do this, we note that in our coordinates, the origin $z = 0$ will
always be included inside the droplet. Moreover, we do not want to
encounter any singularities in $V$ as we move away from the origin
(unless we hit the boundary of the droplet).

 For this, the number of
poles inside the unit circle has to remain unchanged as we smoothly
move away from the origin. Therefore, we can evaluate the poles at
the origin to see whether they are inside or outside the unit
circle. Doing this, we find that $|w_{1,2}|^2 = \epsilon < 1$.
Moreover, $|w_{1,2}|^2/\epsilon^2 = 1/{\epsilon} > 1$. Therefore,
only $w_{1,2}$ will be inside the unit circle. Summing the residues
for finite $z$ we get our final result, \be V = \frac{z(1 +
\epsilon^2) - 2 \epsilon \bar{z}}{1 - \epsilon^2 + \epsilon(z^2  +
\bar{z}^2) - |z|^2(1 + \epsilon^2)} = \bar\partial \log K\;,\ee
where, \be K(z,\bar z) = \frac{1}{1 - \epsilon^2 + \epsilon(z^2  +
\bar{z}^2) - |z|^2(1 + \epsilon^2)}\;.\ee

Note that $K = \infty$ gives the equation of the ellipse with
eccentricity $\epsilon$ \cite{growth}.  Moreover, for $\epsilon = 0$
we recover the usual result, $V = z/(1 - |z|^2)$ which was found in
\cite{sam1, sam2} in the context of open strings on Giant Gravitons.

\subsection{Hypotrochoid}
The Hypotrochoid is shown in figure 3.
\myfig{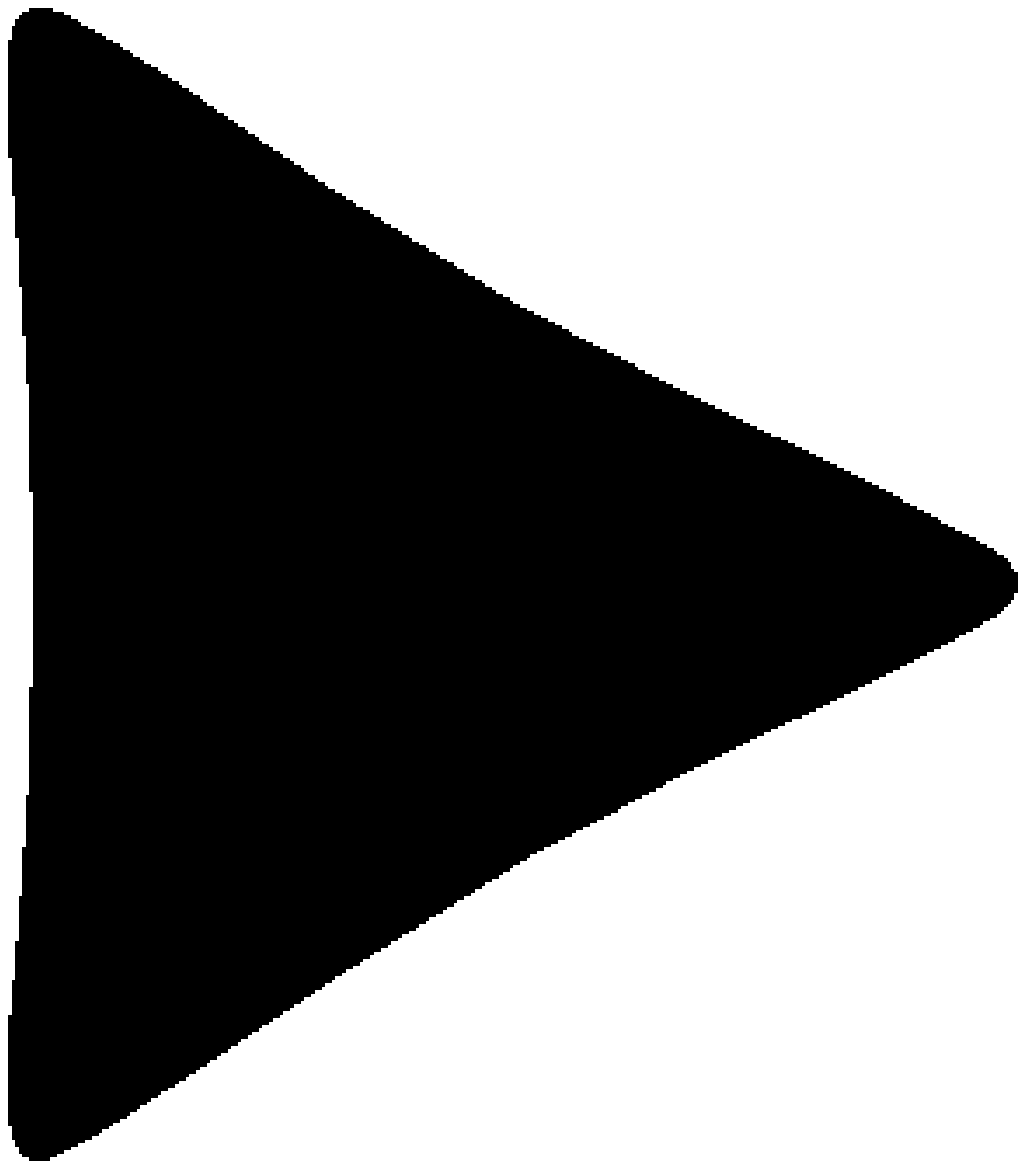}{3}{Hypotrochoid. We have chosen $t_3$ to be real.}
This is another one-parameter family of droplets. It has been also
studied in the context of normal random matrix theory \cite{growth,
sing}. The only non-zero moment is $t_3$. The inverse of the
conformal map is given by, \be \label{hconf} x(w) = r w +
\frac{u}{w^2}\;.\ee The parameters are related by Eqs.
(\ref{areanorm}) and (\ref{moments}), \be 1 = r^2(1 - 2 a^2
r^2)\;,\;\;\; u = 3 t_3 r^2\equiv a r^2\;.\ee Again, one can take
$u$ to be real without lost of generality.

If we vary $a$, we find that the boundary of the droplet develop
singularities at $a > 1/\sqrt{8}$. This kind of singular behavior
has been studied from the point of view of random matrix theory
\cite{sing}. The String Theory interpretation is not clear, but it
would be interesting to understand it. This is, however, outside of
the scope of this paper. In any case, these singularities are of no
concern since, in what follows, we will only expand near small $a$.

We can now calculate $V$ from (\ref{Vint}) as before. One finds that
the integrand contains the inverse of a sixth order polynomial in
$w$. Only three roots are inside the circle. The resulting
expression is quite complicated, so we will only show the first four
orders in $a$:
 \bea \label{Vh} V  &=& \frac{z}{1 - |z|^2} - \frac{\left( z^4 +
{{\bar z}}^2\,\left( 3 - 2\,z\,{\bar z} \right)  \right) \,a}
   {{\left( 1 - |z|^2 \right) }^2} \nonumber \\
   &&+ \frac{\left( 2\,z + z^7 + 3\,{{\bar z}}^5 - 2\,z\,{{\bar z}}^6 \right) \,a^2}
   {{\left( 1 - |z|^2 \right) }^3} \nonumber \\
   &&- \frac{\left( z^4\,\left( 3 + z^6 \right)  + 3\,{{\bar z}}^2 + 3\,{{\bar z}}^8 -
       2\,z\,{{\bar z}}^9 \right) \,a^3}{{\left( 1 - |z|^2 \right)
       }^4} + {\cal O}(a^4)\;.\nonumber \\
\eea In the section VI, we will see how a gauge theory calculation
can reproduce this non-trivial result.

\subsection{Laplacian Growth}

In this section, we will briefly mention how to construct a
one-to-one map from the interior of a simply-connected droplet, to
the interior of the unit disk. We will find a ``quantum" version of
this map in the gauge theory.

Suppose that we normalize the area of the droplet according to, \be
\int_{\cal D} d^2z \rho = \int_{D_R} d^2z \rho\;,\ee where $D_R$ is
circular disk of radius  $R \leq 1$. Our new normalization condition
is then, \be \label{Rdrop} R^2 = \oint_{\partial \cal D}
\frac{dz}{\pi} \bar z\;.\ee  Now consider varying $R$, but keeping
the moments $t_k$ defined in (\ref{moments}) fixed. This process is
known as Laplacian Growth (or shrink, if we decrease $R$)
\cite{matrixreview, growth}. That is, varying $R$ gives
``concentric" droplets.

Using the inverse-conformal map (\ref{invconf}), one can construct
the the desired map from the interior of the unit disk ($D_1$), to
the interior of the droplet: \be L(y) \equiv x( R, w = e^{i\phi})
\in {\cal D}\;,\;\;\; y \equiv R w = R e^{i\phi} \in D_1\;.\ee We
will call $L(y)$ the ``Laplacian Map".

Let us give some explicit examples. For the elliptical droplet, the
map is constructed from the normalization equations (\ref{moments})
and  (\ref{Rdrop}): $R^2 = r^2 - |u|^2$, $u = \epsilon r$. We get,
\be \label{Lmapellipse} L(y) = \frac{1}{\sqrt{1 - \epsilon^2}}
\left( y + \epsilon \bar y\right)\;.\ee

For the Hypotrochoid, one needs to solve, $R^2 = r^2(1 - 2 a^2 r^2)$
and $u = a r^2$. We get, \be\label{Lmaph} L(y) = r \left( y + a r
\bar y^2\right)\;,\ee where \be r = \left(\frac{1 - {\sqrt{1 - 8
a^2|y|^2}}}{4 a^2 |y|^2}\right)^{1/2} \;.\ee Expanding to ${\cal
O}(a^3)$ we get, \be L(y) \approx (1 + a^2 |y|^2) y + a (1 + 2 a^2
|y|^2) \bar y^2 + {\cal O}(a^4)\;.\ee

\section{Gauge Theory Despcription: General Results}
In this section, we show how the reduced sigma model
(\ref{redsigma}) arises directly from the one-loop dilatation
operator of $\cal N$ = 4 SYM in the $SU(2)$ sector. We will do this
by translating the problem to the language of Random Matrix Theory.
Our procedure is very general, and  works for any droplet
configuration, including the case of multiple connected domains.
Moreover, it is straightforward to extend beyond one loop. We will
find that the dilatation operator for any droplet can be interpreted
as a dynamical spin chain. By ``dynamical", we mean that the length
of the chain is not conserved and spins can enter and leave the
lattice.

\subsection{Review and Notation}
The standard operator-state correspondence of ${\cal N} = 4$ SYM
allows us to define a basis for the Hilbert space of states in
$\mathbb{R}\times S^3$ in terms of local operators in
$\mathbb{R}^{1,3}$. The inner product in the Hilbert space is then
mapped to correlation functions of local operators. The inner
product is defined at zero coupling.  If we work only in the scalar
sector of the theory, we can normalize the local operators such that
the propagator takes the form, \be \bra X_i^j(x) \bar{X}_k^l(0)\ket
= \delta_i^l \delta_k^j\;,\ee where $X$ is any of the three complex
scalars of SYM. This is the usual propagator for a gaussian random
matrix model of a single {\it complex} matrix. Thus, we can just
drop the space-time dependence of the operators, and compute all
free-field theory correlation functions using the gaussian matrix
model.

Increasing the coupling away from zero, produces logarithmic
divergences in the OPEs. Extracting these divergences gives the
action of the Dilatation operator. However, the combinatorics can
still be encoded in a simple gaussian matrix model.

In the $SU(2)$ sector, the operators have the general form, \be \psi
\sim \tr(YZYZZ\cdots) \tr(ZYZZ\cdots)\cdots\;.\ee The dilatation
operator acts on these operators as \cite{beisert}, \be\label{dil}
\hat{D} - \hat{J} \equiv \hat H = -\frac{g_{YM}^2}{8\pi^2} \tr[Z,
Y][\partial_Z,
\partial_Y] +  \ldots \ee
The dots indicate higher loop contributions, and we have subtracted
the  R-charge operator ($\hat J$). Moreover, the derivatives have
the property that $(\partial_Z)_i^j Z_k^l = \delta_i^l \delta_k^j$.

An orthonormal basis in this sector is composed of operators such
that, \be \bra\psi_n|\psi_m\ket = \bra \psi_n^*(\bar Y,\bar Z)
\psi_m(Y, Z) \ket = \delta_{n m}\;,\ee where, as usual, \be \bra
{\cal O} \ket =\frac{\int [d^2Y d^2Z] e^{-\tr(|Z|^2 + |Y|^2)} {\cal
O} }{\int [d^2Y d^2Z] e^{-\tr(|Z|^2 + |Y|^2)}}.\ee Here, $[d^2Y
d^2Y]$ is the standard $U(N)$ invariant measure defined by the
metric, $ds^2 = \tr(dY d\bar Y + dZ d\bar Z)$. The matrix elements
of the dilatation operator are defined in the usual way, \be (\hat
H)_{n m} = \bra \psi_n^*(\bar Y,\bar Z) \hat H \psi_m(Y, Z) \ket
\;.\ee

It is now convenient to rescale our fields as $(Y,Z) \rightarrow (Y,
Z)/\sqrt{\hbar}$, where $\hbar = 1/N$. In this way, correlation
functions will be order one in the large $N$ limit ($\hbar
\rightarrow 0$).

The 1/2 BPS sector is given by any {\it holomorphic} wavefunction on
(say) $Z$.  One obviously has, $\hat H \psi(Z) = 0$. In this special
case one can integrate out the off-diagonal components of $Z$. This
is done by expanding $Z = U^\dagger(Z_\text{diag.} + R)U$, where
$Z_\text{diag.}$ is a diagonal matrix, $R$ is an upper-triangular
matrix and $U \in U(N)$. It turns out that the measure of the matrix
model transforms as \cite{matrixreview}, \bea \label{measure}
[d^2Z]e^{-\tr(|Z|^2)/\hbar} &=& \prod_{i < j} \left(d^2R_i^j e^{-
|R_i^j|^2/\hbar} \right)\nonumber
\\
&& \times \prod_{k} d^2 z_k e^{-|z_k|^2/\hbar} \prod_{ l < m } |z_l
- z_m|^2\;.\nonumber \\ \eea Since $\tr((Z_\text{diag.} + R)^n) =
\tr(Z^n)$, the matrix $R$ drops from all holomorphic states. All
correlation functions are then expressed in terms of the eigenvalues
$Z_\text{diag.}$.

We can now  consider a  ``heavy" 1/2 BPS state such as, \be { \psi}
= e^{\tr \; \Omega(Z)/\hbar}\;.\ee The normalization of such a state
is given by the partition function, \be \bra { \psi} | { \psi}\ket
\propto \prod_{i = 1}^N \int d^2 z_i e^{\sum_j W(z_j,\bar{
z}_j)/\hbar + \sum_{i < j} \log |z_i - z_j|^2 }\;,\ee where \be
\label{W} W(z, \bar z) = - |z|^2 + \Omega(z) + \overline{
\Omega(z)}\;.\ee

It is well known that in the ``classical" limit $\hbar \rightarrow
0$, the eigenvalues condense into constant density droplets
\cite{matrixreview, growth} in the complex plane. This is the usual
2D Coulomb gas problem. In fact, the density of the droplets is
$\rho = 1/(\hbar \pi)$.

This matches the String Theory classification of the 1/2 BPS
geometries. However, reconstructing the form of the droplets given
the potential $\Omega(z)$, is a non-trivial inverse problem that has
been the subject of numerous papers. We will not need to go into
these details here, since we will find a new way of doing this.

 Nevertheless, for the case of a single
droplet, the problem simplifies dramatically. One can show
\cite{matrixreview} that the moments introduced in (\ref{moments})
are related to the potential as, \be \label{pot} \Omega(z) = \sum_{k
> 0} t_k z^k \;. \ee This allow us to
reconstruct the operator dual to any single-droplet space-time.

One can also consider potentials that generate multiple droplets.
These potentials have been studied in \cite{multdomain}.
Furthermore, to introduce ``holes" in the droplets one must consider
potentials such as $\Omega(z) \sim  \sum_k a_k \log(z - b_k )$. In
particular, for a single hole in the center of the circular droplet,
one has $\Omega(z) \sim \log z$. This last potential was studied in
\cite{logdrop}. As we will see, our formalism would need to be
slightly modified for this special case, since $\Omega'(z)$ is
singular at $z = 0$. However, it should still be applicable all
other potentials that admit a power law expansion around $z = 0$.

\subsection{Probe String Hilbert Space}
In this section, we will prove some general results that are
applicable for any potential $\Omega(z)$ whose first derivative
$\Omega'(z)$ admits a power-law expansion around $z = 0$. Moreover,
we will always work in the large $N$ limit.

Let us consider a ``probe string" in a 1/2 BPS geometry. This can be
mapped to a periodic lattice, \be\label{probe} |n_1, \ldots, n_L\ket
= \tr(Y\psi_{n_1}(Z) Y  \cdots Y \psi_{n_L}(Z)) e^{\tr\; \Omega(Z)
/\hbar}\;,\ee  with some unknown functions $\psi_n(Z)$. In the large
$N$ limit, the normalization of this state is simply given by, \be
\label{innerp} \bra n_1,\ldots, n_L|n_1', \ldots ,n_L'\ket \approx
\prod_{l = 1}^L \bra \hbar \tr(\overline{\psi_{n_l}( Z)}
\psi_{n_l'}(Z) )\ket\;,\ee where we are considering the generic case
where not all $n_i$ are equal to $n_i'$. At this point, it is
convenient to absorb the factors of $\hbar$ in the normalization of
$\psi_n$. The functions $\psi_n(Z)$ must admit a power law expansion
in $Z$. Moreover, we define $|0\ket \cong \psi_0(Z) \equiv
\sqrt{\hbar} \mathbf{1}$. The alert reader might wonder if $\psi_n$
could also include multiple traces. For example $(\psi)_i^j \sim
\tr(Z^2) Z_i^j + \cdots$. However, one quickly realizes that, since
correlation functions factorize in the large $N$ limit, one can
treat these additional traces as numbers. That is, just replace (for
example) $\tr(Z^2) \rightarrow \bra \tr(Z^2)\ket$.

The correlation functions in (\ref{innerp}) are computed including
the potential $\Omega$, \bea \bra {\cal O}\ket &=& \frac{\int [d^2
Z] e^{\tr W/\hbar} {\cal O}}{\int [d^2 Z] e^{\tr W/\hbar}} \;, \eea
where $W$ is given by (\ref{W}). Thus, our first task is to find an
orthonormal basis such that, \be \label{basis} \bra
\tr(\overline{\psi_{n}( Z)} \psi_{m}(Z) )\ket = \delta_{n m}\;.\ee

For a general potential $\Omega$, this is a highly non-trivial
problem, because the off-diagonal elements of $Z$ cannot be
decoupled. To the best of the author's knowledge, this problem has
not been solved in the Random Matrix Theory literature. Here we will
find a purely algebraic solution to it (in the large $N$ limit).  In
fact, the procedure is very similar to the construction of the
orthogonal polynomials for the Normal Matrix Model \cite{growth}.

Lets start by assuming the existence of the orthonormal basis
(\ref{basis}). One can now define the operator \be \label{Lddef}
(\hat{L}^\dagger)_{n m} = \bra \tr[\overline{\psi_n(Z)} Z
\psi_m(Z)]\ket\;,\ee and its conjugate, \bea\label{Ldefmatrix} (\hat
L)_{n m} &=&
\bra \tr[ \bar Z\overline{\psi_n(Z)}  \psi_m(Z)]\ket \nonumber \\
&=& \bra \tr[ \overline{\psi_n(Z)}( \hbar \partial_Z + \Omega'(Z))
\psi_m(Z)]\ket\;,\eea where the second equality follows by
integrating by parts in the matrix model. Then, one can prove the
following theorem:

{\noindent \it \bf Theorem: } {\it The operators $\hat L^\dagger$
and $\hat L$ obey the algebra,
 \be
  \label{id1} [\hat L, \hat L^\dagger] = \hat P_0\;,
\ee
 \be \label{id2} \hat L = \Omega'(\hat L^\dagger) +
(\hat L^\dagger)^{-1} +
  \sum_{k > 1} v^{(k)} (\hat L^\dagger)^{-k}\;,\ee
where $\hat P_0$ is the projection to $|0\ket$, and $(\hat
L^\dagger)^{-1}$ is defined by left multiplication on $\hat L$.
Moreover, $v^{(k)}$ are constants given by,
 \be \label{vk} v^{(k)} = \hbar \int d^2z \rho(z,\bar z)
z^{k-1} \;\;\;\;\; (k > 1)\;,\ee where $\rho$ is the eigenvalue
distribution of the matrix model.}

{\it Proof:} Lets start by proving (\ref{id1}). For this, we first
need to look at a monomial of $\psi_n$. We have, \bea [\hat L, \hat
L^\dagger] Z^m &=& \hbar ( \partial_Z Z-Z
\partial_Z)Z^m \nonumber
\\
&=& Z^{m} + \sum_{k \geq 1} \hbar \tr(Z^k) Z^{m - k} \nonumber
\\
&&- Z\left( Z^{m-1} + \sum_{k \geq 1} \hbar \tr(Z^k) Z^{m - k -
1}\right) \nonumber \\
&=& \delta_{m,0} + \hbar \sum_{k \geq 1} \delta_{m,k}\tr(Z^k)\;.\eea

In the large $N$ limit, we have \bea\label{corr} \bra \tr(\bar\psi_n
[\hat L, \hat L^\dagger] \sqrt{\hbar} Z^m)\ket &\approx& \bra
\sqrt{\hbar} \tr(\bar\psi_n) \ket\bra \delta_{m,0} \nonumber \\
&&+ \hbar \sum_{k \geq 1} \delta_{m,k}\tr(Z^k) \ket
\nonumber \\
&=& \delta_{n, 0} c_m \;,\eea for some constant $c_m$.

Note that in the Hilbert Space, \be \bra 0| M_m\ket = \hbar \bra
\tr(\mathbf{1} Z^m) \ket = c_m\;. \ee Here $|M_m\ket = \sqrt{\hbar}
Z^m $ is the state corresponding to a monomial, and $|0\ket =
\sqrt{\hbar} \mathbf{1}$. Note that we are normalizing states so
that their inner product is of order one in the large $N$ limit.

Thus, we can write (\ref{corr}) as, \be \label{com} \bra \psi |
[\hat L, \hat L^\dagger] |M_m\ket = \bra \psi | 0\ket\bra 0 |M_m\ket
= \bra \psi_n| \hat P_0 |M_n\ket\;.\ee Since every state
$|\psi_n\ket$ is a linear superposition of monomials $|M_n\ket$, Eq.
(\ref{id1}) follows from (\ref{com}).

Lets us now prove the second identity (\ref{id2}). This one follows
directly form (\ref{Ldefmatrix}). The last two terms in (\ref{id2})
come from the matrix derivative in (\ref{Ldefmatrix}). For a
monomial state, we get, \bea \bra \tr(\bar \psi_n \hbar \partial_Z
\sqrt{\hbar} Z^m )\ket &=&  \sum_{k > 0 } \bra \tr(\bar \psi_n
\sqrt{\hbar} Z^{m-k} ) \hbar \tr(Z^{k-1})\ket
\nonumber \\
&\approx&  \sum_{k > 0 } \bra \tr(\bar \psi_n \sqrt{\hbar} Z^{m-k}
)\ket \bra
\hbar \tr(Z^{k-1})\ket \nonumber \\
&\equiv&  \sum_{k > 0 } \bra \psi_n|(\hat L^\dagger)^{-k}|M_{m}\ket
v^{(k)}\;,\eea where the last line defines the action of the inverse
of $\hat L^\dagger$ and the constants $v^{(k)}$. The proof of
(\ref{id2}) is completed by taking a superposition of monomials.

Since the coefficients $v^{(k)}$ are holomorphic correlation
functions, they can be computed  by an integration over the
eigenvalue distribution as given in (\ref{vk}). For a single droplet
distribution, we can write (\ref{vk}) as, \be \label{vkcont} v^{(k)}
= \oint_{\partial \cal D} \frac{dz}{2\pi i} \bar{z} z^{k - 1}\;.\ee
This completes the proof of the theorem. $ \square $

In practice, one would like to find a particular representation of
the algebra (\ref{id1}) and (\ref{id2}). In this case, we {\it
assume} that the wavefunction $\psi_n(Z)$ is a polynomials of degree
$n$. For the gaussian ensemble ($\Omega = 0$) we have $\psi_n \sim
Z^n$. Clearly, in the general case we will have \be \label{rec} Z
\psi_n(Z) = r_n \psi_{n + 1}(Z) + \sum_{k \geq 0} u_n^{(k)}\psi_{n -
k}(Z)\;.\ee Our goal will be to calculate the coefficients $r_n ,
u_n^{(k)}$. Once this is done, the orthogonal polynomials can be
calculated by iterating (\ref{rec}). Moreover, note that we can
always define $u_n^{(k)}$ so that it gives zero for $k > n$.

We can now translate (\ref{rec})  to the Hilbert Space using
(\ref{Lddef}) and (\ref{Ldefmatrix}). Namely,
 \bea \label{Ldagger}\hat L^\dagger &=&
\hat a^\dagger r_n+ \sum_{k \geq 0}  \hat a^k u_n^{(k)} \;, \\
 \label{L} \hat L &=& r_n \hat a + \sum_{k \geq 0}  \bar{u}_n^{(k)}
(\hat a^\dagger)^k\;,\eea where \be \hat{a}^\dagger|n \ket =
|n+1\ket\;,\;\; \hat a|n\ket = (1 - \delta_{n,0})|n-1\ket\;,\ee and
now $r_n, u_n^{(k)}$ are regarded as diagonal operators. Note that
$\hat a, \hat a^\dagger$ are the Cuntz oscillators used in
\cite{sam1, sam2}.  Moreover, one can always take $r_n$ to be real
without lost of generality.

The inverse operator $(\hat L^\dagger)^{-1}$ is defined by left
multiplication on $\hat L$: $(\hat L^\dagger)^{-1} \hat L^\dagger =
1$. Explicitly, \bea(\hat L^\dagger)^{-1} &=& [(r_{n-1} + \sum_{k
\geq
0} \hat a^{k +1} u_n^{(k)} )\hat a^\dagger]^{-1} \nonumber \\
&=&  \hat a \frac{1}{r_{n-1} + \sum_{k \geq 0} \hat a^{k +1}
u_n^{(k)}}\;,\eea where the last expression is defined by its power
expansion in $\hat a$.

The coefficients $r_n$ and $u_n^{(k)}$ in (\ref{Ldagger}) and
(\ref{L}), are completely determined by the algebra (\ref{id1}) and
(\ref{id2}). In particular, after some work, one can show that the
diagonal part of (\ref{id1}) reads, \be \label{diag} 1 = r_n^2 -
\sum_{k \geq 1} \sum_{p = 0}^n |u_{n + k}^{(k + p)}|^2\;.\ee

{\it Summary}: This section has been quite technical, so let us
summarize the main results. The Hilbert space for a probe string in
a generic 1/2 BPS geometry can be mapped to a periodic lattice as in
(\ref{probe}). This basis is endowed with the algebra (\ref{id1})
and (\ref{id2}). One then assumes that the wavefunctions $\psi_n$
are polynomials. Their explicit form is defined by the recursion
relation (\ref{rec}). The coefficients of the recursion relation are
completely determined by the algebra (\ref{id1}) and (\ref{id2}), by
using the representation (\ref{Ldagger}) and (\ref{L}).

\subsection{Hamiltonian}

In this section, we will find the action of the Hamiltonian
(\ref{dil}) in the probe string basis studied in the previous
section.  We will only work at one-loop. First, let us study the
action on a particular $Y$ on the lattice, \bea &&\hat H |\cdots,
n_l, n_{l+1},\cdots\ket =-\frac{g_{YM}^2 N}{8\pi^2} \tr[Z, Y][\hbar
\partial_Z,
\partial_Y]\nonumber \\
&&\times \tr(\cdots \psi_{n_l} Y \psi_{n_{l+1}} \cdots ) e^{\tr
\; \Omega(Z)/\hbar}\nonumber \\
&&= -\frac{g_{YM}^2 N}{8\pi^2} \left[ \tr(\cdots \psi_{n_l} [Z, Y]
\hbar \overleftrightarrow{\partial_Z} \psi_{n_{l+1}} \cdots )
\right.
\nonumber \\
&& \left. -\tr(\cdots \psi_{n_l}  \hbar
\overleftrightarrow{\partial_Z} [Z, Y] \psi_{n_{l+1}} \cdots )
\right] \nonumber \\
&&\times e^{\tr \;\Omega(Z)/\hbar}\;.\eea The double arrow means
that the derivative acts on both sides but always {\it excluding}
the $Z$ in the commutator. Note that the derivative will also act on
the potential $\Omega(Z)$.

In the planar limit, one can show that multiple traces in $Y$ are
still suppressed. Therefore, one must not allow the derivative to
act beyond its own site. Then, it is easy to show that the action of
the Hamiltonian has a remarkably simple form in terms of the
operators (\ref{Ldagger}) and (\ref{L}): \be \label{H}\hat H =
\lambda \sum_{l = 1}^L (\hat L^\dagger_l - \hat L^\dagger_{l +
1})(\hat L_l - \hat L_{l + 1}) \;,\ee where periodic boundary
conditions are understood.

The  form of the Hamiltonian (\ref{H}), is very similar to the
bosonized version of the ferromagnetic Heisenberg spin chain
introduced in \cite{sam1, sam2}. In the general case, however, one
has a complicated canonical structure given by (\ref{id1}) and
(\ref{id2}).

To get the usual $XXX_{1/2}$ spin chain, one takes the simplest case
of $\Omega = 0$. This choice gives the disk distribution of
eigenvalues. Using (\ref{vkcont}) one can easily prove $v^{(k)} =
0$. Thus, the algebra (\ref{id1}) and (\ref{id2}) reduces to, \be
[\hat L, \hat L^\dagger] = \hat P_0\;, \;\;\;\; \hat L \hat
L^\dagger = 1\;.\ee In other words: \be \label{H0}\left.\hat H
\right|_{\Omega = 0} = \lambda\sum_{l = 1}^L (\hat a^\dagger_l -
\hat a^\dagger_{l + 1})(\hat a_l - \hat a_{l + 1}) \;.\ee

To translate this to the spin chain language, one makes the
following identification \cite{sam1, sam2}, \be \label{spin} |n_1,
n_2,\ldots ,n_L\ket = |\uparrow \underbrace{\downarrow
\downarrow\cdots \downarrow}_{n_1} \uparrow \underbrace{\downarrow
\downarrow\cdots \downarrow}_{n_2} \uparrow \cdots \uparrow
\underbrace{\downarrow \downarrow\cdots \downarrow}_{n_L} \ket
\;.\ee In the spin chain basis, the Hamiltonian (\ref{H0}) can be
written as \be \left.\hat H \right|_{\Omega = 0} =  \lambda \sum_{l
= 1}^J \left(1 - 4 \vec{S}_l \cdot \vec{S}_{l+1}\right)\;,\ee where
$J = L + \sum_{l = 1}^L n_l$. This is indeed the spin chain that was
originally discovered in the context of the AdS/CFT correspondece in
\cite{zarembo}.

If we now consider the general case with $\Omega \neq 0$, one can
easily see from (\ref{Ldagger}), (\ref{L})  and (\ref{H}), that the
number of ``spin downs" in (\ref{spin}) will not be conserved. In
other words, the length of the spin chain is not preserved. In this
case, the best interpretation for the Hamiltonian (\ref{H}), is as a
bosonic lattice with $n_l$ bosons at each site. Moreover, the bosons
can leave or enter the lattice at any site. This kind of dynamical
lattice was first found in \cite{sam1, sam2} in the study of Giant
Gravitons. Dynamical lattices have also been studied (in a different
context) in \cite{dynamicspin, condmatter}.

\subsection{Coherent States and Space-Time Metric}

If we want to gain insight into the classical limit of the
Hamiltonian (\ref{H}), one must construct coherent states of the
operator $\hat L$. Making the general ansatz for the coherent
states: \be |z\ket = \sum_{n = 0}^\infty f_n(z)|n\ket\;,\ee it is
easy to show that one needs, \be z f_n(z) = r_{n} f_{n+1}(z) +
\sum_{k\geq0} \bar u_n^{(k)} f_{n - k}(z)\;.\ee So we see that $f_n$
are really the complex conjugate of the wavefunctions (\ref{rec}):
\be f_n(z)  = \psi_n^*(z)\;.\ee Moreover, the range of the
coordinate $z$ will be given by the normalization condition,
\be\label{norm} 0 \leq \bra z| z\ket < \infty\;.\ee This condition
will give us the shape of the droplet.

Now, we have not proved that these coherent states are
(over)complete. This is, in general, very difficult. We remind the
reader that, even for the simple case of the gaussian ensemble with
$f_n = z^n$, the completeness relation requires to introduce a very
special measure in terms of the so-called ``Jackson Integral"
\cite{sam1,sam2}. This turns out to be irrelevant in our case, since
we will only use the coherent states in the classical limit, where
the measure of the path integral can be ignored. So from now on, we
just {\it assume} that a measure exist, such that $1  = \int
d\mu(z,\bar z) |z\ket\bra z|$.

In this case, one can always write the classical action for a
general coherent state $|CS\ket$ as \cite{coh}, \be S = \int
d\tau\left( i\bra CS|\partial_\tau|CS\ket  - \bra CS|\hat H|CS\ket
\right)\;.\ee In our case this reduces to, \bea \label{cslag}S &=&
\int d\tau \sum_{l = 1}^L \left( i \bra z_l|\partial_\tau|z_l\ket -
\lambda
|z_l - z_{l + 1}|^2 \right)\nonumber \\
&\rightarrow& -  L \int d\tau \int_0^1 d\sigma \left( \frac{i}{2} V
\dot{\bar z} - \frac{i}{2} \bar V \dot z + \frac{\lambda}{L^2}
|z'|^2\right)\;, \eea where \be\label{Vstring} V = \bar \partial
\log\left(\sum_{n = 0}^\infty |\psi_n(\bar z)|^2\right)\;.\ee

Therefore, we find that in the thermodynamic limit $L\rightarrow
\infty$, the coherent state action of our quantum lattice has the
same form as the String Theory result (\ref{redsigma}). The extra
constant in (\ref{redsigma}) can be obtained if we add the R-charge
operator $\hat J_Y = \tr(Y\partial_Y)$.

 One can
readily identify the function $V$ defined in (\ref{Vstring}) with
the one-form (\ref{V}). Moreover, from (\ref{Kahler}) we can also
reconstruct the remaining function in the metric, \be
\label{hstring} h^4 =
\partial \bar \partial \log\left(\sum_{n = 0}^\infty |\psi_n(\bar z)|^2\right)\;.\ee

The reader can easily verify that for the circular droplet, with
$\psi_n = z^n$, both (\ref{Vstring}) and (\ref{hstring}) reduce to
the familiar results, \be V = \frac{z}{1 - |z|^2}\;,\;\;\; h^4 =
\frac{1}{(1 - |z|^2)^2}\;.\ee

In practice, in order to perform the sum over the orthogonal
polynomials, one needs to find their generating function: \be
\label{gen} G(z,x) = \sum_{n = 0}^\infty f_n(z) x^n\;.\ee For a
general droplet, this is a difficult object to construct. However,
we will see that one can set up a systematic expansion around the
circular droplet.

\section{Example 1: Elliptical Droplet}

This is the simplest droplet next to the circular one. It is
generated by the simple potential $\Omega(z) = t_2 z^2$. From
(\ref{id2}) it is easy to see that $\hat L^\dagger$ will truncate at
${\cal O}(\hat a)$: \be \hat L^\dagger = \hat a^\dagger r_n +
u^{(0)} + \hat a u^{(1)}_n \;.\ee

Collecting terms of ${\cal O}(\hat a^\dagger)$ and ${\cal O}(1)$ in
(\ref{id2}) one finds that, \be \frac{\bar u^{(0)}_n}{u_n^{(0)}} =
\epsilon\;,\;\;\; \frac{\bar u_n^{(1)}}{r_{n+1}} = \epsilon\;,\ee
where $\epsilon = 2 t_2$. Since we want to be able to consider the
case $\epsilon < 1$, it follows that $u_n^{(0)} = 0$. Moreover, from
(\ref{diag}) it follows that, \be 1 = r_n^2 - |u_{n+1}|^2\;.\ee
Thus, we find \be \label{esol} r_n = \frac{1}{\sqrt{1 - \epsilon^2}}
\;,\;\;\; u_n = \frac{\epsilon}{\sqrt{1 - \epsilon^2}}\;,\ee where
we have chosen $\epsilon$ real without lost of generality.

One can explicitly check that the other non-diagonal terms in
(\ref{id2}) are indeed satisfied. For example, at ${\cal O}(\hat a)$
one gets the equation, \be {\hat a} (1 - r_n^2 + \epsilon \;u_n r_n)
= 0\;.\ee
 which is easily seen to be satisfied if we use (\ref{esol}).

 We can now find the generating function of the polynomials by using
 their recursion relation:
 \be z G(z,x) = \sum_{n = 0}^\infty \left( \frac{1}{\sqrt{1 -
 \epsilon^2} } f_{n + 1}+ \frac{\epsilon}{\sqrt{1 - \epsilon^2}}
 f_{n - 1} \right) x^n\;,\ee
 from where we find,
 \be G(z ,x) = \frac{1}{\epsilon x^2 - \sqrt{1 - \epsilon^2} x z +
 1}\;.\ee This is, in fact, the generating function of the Chebyshev
 Polynomials of the Second Kind.

 The sum over the orthogonal polynomials can be calculated, in
 general, from
 \be \label{sumint} \sum_{n = 0}^\infty |f_n(z)|^2 = \oint_{S^1} \frac{dx}{2\pi i} \frac{1}{x} G(z, x)
 \bar G(\bar z , 1/x)\;.\ee

 We can calculate this integral explicitly using the Residue
 Theorem. Just like in the String Theory calculations, one can find out which pole is
inside the unit circle by slowly moving away from the point $z = 0$.
Alternatively, one can require that the solution should be
continuously connected with the case $\epsilon = 0$. In any case,
one finds that the only poles inside the unit circle are the roots
of the quadratic equation, \be \epsilon - \sqrt{1 - \epsilon^2} \bar
z x + x^2 = 0\;.\ee

 The sum of the residues gives,
 \be \sum_{n = 0}^\infty |f_n(z)|^2 = \frac{1}{1 - \epsilon^2 + \epsilon(z^2  +
\bar{z}^2) - |z|^2(1 + \epsilon^2)}\;.\ee This is exactly what we
found in the String Theory calculation! From this ``Kahler
Potential" we can reconstruct the whole metric using (\ref{Vstring})
and (\ref{hstring}). We clearly see that the shape of the droplet
coincides with the range of $z$ that gives normalizable coherent
states (c.f. (\ref{norm})).

We want to point out that the elliptical droplet is rather special.
One can, in fact, find the explicit form of the orthogonal
polynomials by a direct matrix model calculation. Using a simple
shift in the integration variables in the matrix model partition
function, one can show, \be e^{2 \tr(|X|^2)/\epsilon} = \bra
e^{\tr(-X^2 + \alpha Z X)} e^{\tr(-\bar X^2 + \alpha \bar Z \bar
X)}\ket\;,\ee where $\alpha = \sqrt{2(1 - \epsilon^2)/\epsilon}$.

The orthogonal matrix polynomials are given by, \be \psi_n(Z) =
\sqrt{\frac{\epsilon}{2 N^{n + 1}}} \, (\partial_X)^n \left.
e^{\tr(-X^2 + \alpha Z X)} \right|_{X = 0}\;.\ee In the large $N$
limit, the derivatives $\partial_X$ must act in such a way as to
avoid the creation of multiple traces in $\partial_X$. Moreover, we
need to remember that any multiple traces in $Z$ must be replaced by
their expectation value. Then, it is easy to show that these are the
same polynomials as the ones found above.

Let us close this section with an interesting observation regarding
the operator $\hat L$. Let us identify the operators $\hat a$ and
$\hat a^\dagger$ as coordinates on a unit disk: \be \hat a \sim
y\;,\;\;\; \hat a^\dagger \sim \bar{y}\;,\;\;\; y\in D_1\;.\ee This
identification actually follows from the coherent states of the
operator $\hat a$ which are normalized only inside the unit disk
\cite{sam1,sam2}.

With this identification, the operator $\hat L$ can be seen to give
the classical map (\ref{Lmapellipse}) from $D_1$ to the interior of
the elliptical droplet. We will find that, in general, $\hat L$
gives a ``quantum" version of this map (for a single droplet).

\section{Example 2: Hypotrochoid}
This droplet is another example of a one-parameter family of
potentials given by $\Omega(z) = t_3 z^3$. From Eq. (\ref{id2}), one
can see that the operator $\hat L$ must truncate at ${\cal O}((\hat
a^\dagger)^2)$. Thus, we are left with the ansatz, \be \hat L = r_n
\hat a + \bar u^{(0)}_n + \bar u^{(1)}_n \hat a^\dagger + \bar
u_n^{(2)} (\hat a^\dagger)^2\;.\ee

Solving the recursion relations with this general ansatz is very
cumbersome. However, one can make the simplifying assumption that
 $u_n^{(0)} = u_n^{(1)} = 0$. The motivation for this, is the
 similarity between the operator $\hat L$ and the conformal map
 (\ref{hconf}) (for a single droplet). This same kind of assumptions are used in the
 context of the classical orthogonal polynomials \cite{growth}. We will
 latter see that this is a self-consistent assumption.

 With this simplification, the ${\cal O}((\hat a^\dagger)^2)$ term from
 (\ref{id2}) gives the relation,
 \be \label{u2} \bar u_n^{(2)} = a r_{n-1} r_{n-2}\;,\ee
 where we have defined $a \equiv 3 t_3$ which we take to be real
 without lost of generality. Finally, Eq. (\ref{diag}) gives,
 \be 1 = r_n^2 - |u_{n+1}^{(2)}|^2 ( 1 - \delta_{n , 0}) -
 |u_{n+2}^{(2)}|^2\;.\ee
 Therefore, we get a closed equation for $r_n$:
 \be 1  = r_n^2[1 - a^2 (r_{n-1}^2 (1 - \delta_{n,0}) +
 r_{n+1}^2)]\;.\ee
 One can solve this order by order in $a$. We have,
 \be \label{rapprox} r_{n}^2 \approx 1 + a^2(2 - \delta_{n,0}) + {\cal O}(a^4)\;.\ee
Using this result in (\ref{u2}) we obtain, \be u_n^{(2)} \approx a +
a^3 \left(2  - \frac{1}{2} \delta_{n-2,0}\right) + {\cal
O}(a^5)\;.\ee Note that we have discarded the term with
$\delta_{(n-1),0}$ since we always have $n \geq 2$ for this
operator.

These results allow us to reconstruct the recurrence relation for
the coherent states. Namely, \bea  \label{rech} z f_n(z)
&=&\left[\left(1 + \frac{1}{2} a^2 \right)  + \frac{1}{2}a^2\left(1
- \delta_{n,0}\right)\right]
f_{n+1}(z)\nonumber \\
&& + \left[ a\left( 1 +  \frac{3}{2} a^2\right) +\frac{1}{2} a^3
\left( 1-
\delta_{n,2}\right)\right]f_{n-2}(z) \nonumber \\
&& + {\cal O}(a^4)\;.\eea

This recurrence relation allow us to solve for the generating
function to ${\cal O}(a^3)$ accuracy. We have, \bea z G(z, x) &=&
\frac{1}{x} \left(1 + \frac{1}{2} a^2 \right) \left(G(z,x) -
1\right) \nonumber \\
&& + \frac{1}{2x} a^2\left(G(z,x) - 1 - f_1(z) x\right) \nonumber
\\
&& + a\left( 1 +  \frac{3}{2} a^2\right) x^2 G(z,x) + \frac{1}{2}
a^3 x^2  \left(G(z,x) - 1\right) \nonumber \\
&& + {\cal O}(a^4)\;.\eea

The polynomial $f_1(z)$ can be constructed explicitly from the
recursion formula (\ref{rech}), \be f_1(z) = \frac{z}{1 +
\frac{1}{2} a^2}\;.\ee Note that this formula should only be
understood as an approximation which is good to order $a^3$.

The generating function can now be explicitly written as, \be
\label{Gh} G(z,x) \approx \frac{\left( 2 + a^2 \right) \,\left( 2 +
a^2\,\left( 2 + a\,x^3 \right)  \right)  + 2\,a^2\,x\,z}
  {2\,\left( 2 + a^2 \right) \,\left( 1 + a\,\left( a + x^3 + 2\,a^2\,x^3 \right)  - x\,z \right)
  }\;.\ee
In this last result, we have chosen not to expand in powers of $a$,
since the singularity of $G$ is very important for the next
calculation. However, one must keep in mind that the final result is
only valid up to corrections of order $a^4$.

To calculate the sum of the orthogonal polynomials, we use
(\ref{sumint}). Following a similar procedure as with the elliptical
droplet, one finds that the integrand has simple poles inside the
unit circle for  $x = 0$, and at the roots of, \be a + 2 a^3  - \bar
z x^2 + (1 + a^3)x^3 = 0\;.\ee The final sum over the residues is
quite complicated, but the one form $V$ (c.f.  Eq. (\ref{Vstring}))
simplifies a bit. The answer turns out to be exactly the one found
in the String Theory calculation, Eq. (\ref{Vh})!

Let us now return to the interpretation of $\hat L$ as the Laplacian
map. Using our results for $r_n$ and $u_n^{(2)}$ one can write $\hat
L$ as, \bea \label{Lhgauge} \hat L &\approx& \left(1 + \frac{1}{2}
a^2 \hat a^\dagger \hat a\right)\hat a + \hat a \left(\frac{1}{2}
a^2 \hat
a^\dagger \hat a\right) \nonumber \\
&&+ a \left(1 +
\frac{3}{2} a^2 \hat a^\dagger \hat a \right) (\hat a^\dagger)^2 \nonumber \\
&&+ \frac{1}{2} a^3 (\hat a^\dagger)^2 \hat a^\dagger \hat a + {\cal
O}(a^4)\;. \eea If we interpret $\hat a \sim y$ and $\hat a^\dagger
\sim \bar y$ as coordinates on $D_1$, and we ignore their ordering,
one obtains precisely the Laplacian map (\ref{Lmaph}). Note,
however, that some of the terms  in (\ref{Lhgauge})  are trivial
(e.g. $\hat a \hat a^\dagger \hat a = \hat a$). Nevertheless, they
are important for the interpretation as a Laplacian Map.

To finish this section, let us check that our initial assumption,
$u_n^{(0)} =  0 = u_n^{(1)}$, is consistent. For this, we can just
check explicitly the orthonormality of some of the polynomials
$\psi_n(Z)$. Let us consider some examples.

The first few polynomials that follow from the generating function
(\ref{Gh}) are (remember that $f_n(z) =  \psi_n^*(z)$, and we are
taking $a$ to be real), \bea \psi_0(Z) &=& 1\;,\\
\psi_1(Z) &=& Z\left(1 - \frac{1}{2} a^2\right) + {\cal O}(a^4)\;,\\
\psi_2(Z) &=& Z^2\left(1 - \frac{3}{2} a^2\right) + {\cal O}(a^4)\;,\\
\psi_3(Z) &=& Z^3\left(1 - \frac{5}{2} a^2\right)  - a\left(1 +
\frac{1}{2} a^2\right) + {\cal O}(a^4)\;,\\
\psi_4(Z) &=& Z^4\left( 1 - \frac{7}{2} a^2\right) - 2 a Z + {\cal
O}(a^4)\;. \eea

The simplest orthogonality condition to check is $\bra 0|n\ket = 0$,
which in matrix form reads, \bea \bra 0 | n\ket &=& \bra \hbar
\tr(\overline{\psi_0(Z)} \psi_n(Z)\ket \nonumber \\
& =& \int_{\cal D} \frac{d^2z}{\pi} \psi_n(z)  \nonumber \\
&=& \oint_{\partial \cal D} \frac{dz}{2\pi i} \bar z
\psi_n(z)\nonumber \\
&=& \oint_{S^1} \frac{dw}{2\pi i} x'(w) \bar{x}(w^{-1}) \psi_n(x(w))
\;,\eea where $x(w)$ is given in (\ref{hconf}). One can check
explicitly that this expression is zero since there are no simple
poles inside $S^1$.

Now, let us consider some non-trivial cases, where the off-diagonal
elements of $Z$ do not drop out. As an example, we can calculate the
normalization of $\psi_1(Z)$. Using the measure change
(\ref{measure}), we get \bea \bra 1| 1\ket &=&  (1 - a^2 + {\cal
O}(a^4))\bra \hbar \tr(|Z|^2)\ket \nonumber \\
&=&(1 - a^2 + {\cal O}(a^4))\left( \hbar \bra
\tr(|Z_\text{diag.}|^2)\ket + \hbar \bra \tr(\bar{R}R)\ket \right)
\nonumber\\
 & \approx & (1 - a^2 + {\cal O}(a^4)) \left(\frac{1}{2}
\oint_{\partial \cal D} \frac{d^2z}{2\pi i } \bar{z}^2 z +  \hbar^2
 \frac{1}{2} N^2\right) \nonumber \\
 &=&(1 - a^2 + {\cal O}(a^4))\left( 1 + a^2 + {\cal O}(a^4) \right)\nonumber \\
 &=& 1 + {\cal O}(a^4)\;.  \eea

Finally, let us check the highly non-trivial result $\bra 1 | 4\ket
= 0$. For this, we can use the fact that, after integrating out $R$,
one gets \be \bra \tr(\overline{Z} Z^4)\ket_R = \sum_i \bar{z}_i
z_i^4 + \sum_i z_i^3+ \hbar \sum_{i, j} z_i z_j^2 - 2 \hbar \sum_i
z_i^3\;,\ee where $z_i$ are the elements of $Z_\text{diag.}$.
Therefore, in the large $N$ limit, we can write, \bea \bra 1 | 4
\ket &=& (1 - 4 a^2 + {\cal O}(a^4))
\left[\frac{1}{2}\oint_{\partial \cal D} \frac{dz}{2 \pi i}
\bar{z}^2 z^4+ \oint_{\partial \cal D} \frac{dz}{2 \pi i} \bar{z}
z^3
 \right.\nonumber \\
&&\left. +  \left(\oint_{\partial \cal D} \frac{dz}{2 \pi i} \bar{z}
z \right) \left(\oint_{\partial \cal D} \frac{dz}{2 \pi i} \bar{z}
z^2 \right) \right] \nonumber \\
&& +\left(-2 a + a^3 + {\cal O}(a^4)\right)(1 + a^2 +
{\cal O}(a^4))\nonumber \\
&=& (1 - 4 a^2 + {\cal O}(a^4)) (2 a + 9 a^3 + {\cal O}(a^4))
\nonumber \\
&&+\left(-2 a + a^3 + {\cal O}(a^4)\right)(1 + a^2 + {\cal
O}(a^4))\nonumber \\
&=& {\cal O}(a^4)\;. \eea

Therefore, we conclude that our initial ansatz, $u_n^{(0)} =  0 =
u_n^{(1)}$, was indeed correct.

\section{Discussion and Future Directions}
In this paper, we have derived the one-loop Hamiltonian for an
$SU(2)$ probe string on a generic 1/2 BPS background, defined by the
CFT operator $\psi \sim \exp(\tr\; \Omega(Z) /\hbar)$. We found that
the Hamiltonian can be written as (\ref{H}), where the operators
$\hat L$, $\hat L^\dagger$ obey the algebra (\ref{id1}) and
(\ref{id2}). We also found a representation of this algebra in terms
of the Cuntz oscillators. In this basis, the Hamiltonian can be
interpreted either as a dynamical spin chain, or as a bosonic
lattice where the total number of bosons is not conserved.

We found that, in general, the full metric on the reduced space of
the probe string, can be calculated from the coherent states of
$\hat L$ (c.f. (\ref{Vstring}) and (\ref{hstring})). Finally, we
studied two special potentials ($\Omega$) dual to the elliptical
droplet and the Hypotrochoid. We found perfect agreement with the
String Theory results.

\subsection{Generalizations}
Let us now discuss the range of validity of our calculations, and
the possible generalizations of our results. First of all, note that
the algebra (\ref{id1}) and  (\ref{id2}) is valid for {\it any}
potential $\Omega(z)$, whose first derivative has a power-law
expansion around $z = 0$. Moreover, these equations are valid
regardless of the form of $\psi_n(Z)$.  Therefore, the assumption
that $\psi_n$ are polynomials does not affect this result. This
assumption is only used, when we want to find an explicit
representation of the algebra. Thus, in general, the ``droplets" are
defined by the requirement that the coherent state of $\hat L$ is
normalizable (regardless of the form of $\psi_n$).

It would be very interesting to study the case of multiple domains,
and non-simply connected domains. As we mentioned before, this last
case can be a bit tricky if $\Omega'(z)$ is not well defined. Thus,
the case of $\Omega = \log z$ requires special treatment. We expect
that this potential generates a hole in the center of the circular
droplet. In fact, this potential has been studied for normal
matrices in \cite{logdrop}.

 Nevertheless, one can
also consider creating holes in other parts of the droplets using
$\Omega(z) \sim \sum_k c_k \log(z - z_k)$. The first derivative of
this potential is well defined. However, the actual representation
of the algebra can be quite complicated \footnote{See \cite{growth}
for an example of classical orthogonal polynomials with a
logarithmic potential.}. For multiple connected domains, one can
look at the potentials studied in \cite{multdomain}.

The case of multiple domains is very interesting from the point of
view of the  Hamiltonian (\ref{H}). This is because we expect that,
somehow, the Hilbert space should be a direct sum over sub-Hilbert
spaces for each droplet. This is very analogous to what happens with
Topological Field Theories. It would be very interesting to
understand how this happens.

For non-simply connected domains, one can wrap strings around
non-contractible cycles. Therefore, one expects that the
corresponding Hamiltonian has topologically stable solitonic states.
It would be interesting to understand more about these states.

Finally, it would be interesting to extend these techniques to other
sectors in the gauge theory. In particular, one expects that the
$SL(2)$ sector  describes strings propagating outside the droplets,
but still at $y = 0$. String in this sector (on $AdS_5\times S^5$)
have been studied recently in \cite{Park,Bellucci,Diego}. Moreover,
one can leave the $y = 0$ plane by considering bigger sectors such
as $SU(1,1|2)$ \cite{su112}.

\subsection{Integrability}
In recent years, there has been great interest in using
integrability to test/prove the AdS/CFT correspondence. This
interest was sparkled by the discovery of one-loop integrability for
single trace scalar operators \cite{zarembo}. Integrability in the
gauge theory has been argued to persist at higher loops and an
all-loop guess for the Bethe ansatz has been presented in
\cite{beisert1, beisert2}. This has also been accompanied by a
similar guess for the presumed quantum Bethe ansatz for the dual
string theory \cite{Arutyonov}.

However, all these developments are only relevant for single-trace
operators. That is, a probe string on $AdS_5 \times S^5$. It is
doubtful that full integrability will be preserved for a generic 1/2
BPS background. However, it could happen that integrability is still
present for some reduced sub-sectors. The algebra found in this
paper is indeed very suggestive.

If integrability is indeed present around 1/2 BPS geometries, it
must be realized in a very exotic way. This is because, as we have
seen, the lattice models found in this paper do not preserve the
number of ``particles". Thus, the usual Bethe Ansatz is totally
useless. Perhaps one could directly construct conserved charges
using the algebra (\ref{id1}) and (\ref{id2}).

Another possible source of integrable structures might come from the
matrix model itself. It is well known that matrix models show
integrable Toda hierarchies (see \cite{interev} for a review). In
this context, the ``times" of the Toda hierarchy are identified with
the moments of the droplet $t_k$. Perhaps such a structure, if still
present in our model, could be used to ``adiabatically" evolve the
spectrum of a single droplet away from the circular one. Whether
this is possible or not, remains to be seen.

\subsection{Probing Black Hole States}
One of the greatest challenges of the AdS/CFT correspondence, is to
give us a better understanding of black hole physics. In this
context, even if we identify the operator dual to a black hole
microstate, we need a way to measure the resulting metric. The tools
developed in this paper, can be considered as a first step in this
direction. In the $SU(2)$ sector, one can start with the so-called
``superstar" configuration \cite{superstar}. This is a singular 1/2
BPS geometry. It is the extremal limit of a one R-charge black hole
\cite{gubser}.

This geometry must be interpreted as a limit of a very excited 1/2
BPS state. The limit correspond to exciting a large triangular Young
Tableux as advocated in \cite{vijay}. The precise form of the
resulting operator is unknown. However, the dual ``droplet" must
consist of a series of concentric rings. In the limit where these
rings become very thin and closely spaced, we will get a circular
droplet, but with $\rho < 1/\pi$. Since the operator for this state
is holomorphic, the off-diagonal elements of the matrix $Z$ will
drop out. In the eigenvalue basis, the operator should be described
by a Fractional Quantum Hall state \cite{LLL}. It would be
interesting to probe geometries like this using our formalism. This
can give us a better understanding of the emergence of singularities
in AdS/CFT.

However, the really interesting story starts to develop when we move
away from extremality. According to the dictionary in \cite{gubser},
this amount to adding some excitation to the superstar operator such
that we create an anomalous dimension. The size of the anomalous
dimension is, in fact, directly related to the non-extremality
parameter of the resulting Black Hole. Within the $SU(2)$ sector,
one can imagine adding a few $Y$ excitations. These excitations will
produce a finite anomalous dimension. In fact, adding $Y$ fields
correspond to exciting open string on Giant Gravitons
\cite{davidemergent}. This is analogous to the mechanism advocated
in \cite{gubser} to explain the entropy of R-charged black holes.

From the point of view of the matrix model, adding  $Y$ fields will
produce a dramatic change in the norm of the state. This is because
the  off-diagonal elements of $Z$ will no longer drop out. It is
then desirable to learn how to probe such an excited state. Do we
really create a Black Hole? Can we measure the resulting metric?
These questions will be left for future works \cite{work}.

\acknowledgments I would like to thank David Berenstein, Joe
Polchinski, Gary Horowitz, Don Marolf, Diego Correa and Sean
Hartnoll for interesting discussions. This work has been supported,
in part, by an NSF Graduate Research Fellowship.

\end{document}